\documentclass[showpacs,aps,graphicx]{revtex4}
\usepackage{amsmath}
\usepackage{graphicx}

\begin{document}

%
%
%
%

\title{Universal quantum gates on electron-spin qubits with quantum dots inside single-side optical
microcavities\footnote{Published in Opt. Express \textbf{22},
593-607 (2014)}}

\author{Hai-Rui Wei$^1$ and Fu-Guo Deng$^{1,2,}$\footnote{Corresponding author:  fgdeng@bnu.edu.cn}}

\address{$^1$Department of Physics, Applied Optics Beijing Area Major Laboratory,
Beijing Normal University, Beijing 100875, China\\
$^2$State key Laboratory of Networking and Switching Technology,
Beijing University of Posts and Telecommunications, Beijing 100876,
China}

\begin{abstract}
We present some compact quantum circuits for  a deterministic
quantum computing on electron-spin qubits assisted by quantum dots
inside single-side optical microcavities, including the CNOT,
Toffoli, and Fredkin gates. They are constructed by exploiting the
giant optical Faraday rotation induced by a single-electron spin in
a quantum dot  inside a single-side optical microcavity as a result
of cavity quantum electrodynamics.  Our universal quantum gates have
some advantages. First, all the gates  are accomplished with a
success probability of 100\% in principle. Second, our schemes
require no additional electron-spin qubits and they are achieved by
some input-output processes of a single photon. Third, our circuits
for these gates are simple and economic. Moreover, our devices for
these gates work in both the weak coupling and the strong coupling
regimes, and they are feasible in experiment.
\end{abstract}



\maketitle

%
%


\section{Introduction}\label{sec1}

In quantum computing, a quantum algorithm is usually realized by a
sequence of quantum gates \cite{book}. Constructing compact quantum
gates  is crucial for building a quantum computer. It has been
proven that any quantum entangling gate supplementing with
single-qubit gates can implement a universal quantum computing
\cite{uni}.  The controlled-not (CNOT) gate is a universal two-qubit
gate and it attracts much attention. As for multi-qubit quantum
systems, attention was mainly focused on the three-qubit Toffoli and
Fredkin gates as they can be used to implement any multi-qubit
quantum computing with Hadamard gates \cite{Toffoli,Fredkin}.

Up to now, many important  proposals have been proposed for
physically implementing  quantum gates
\cite{longpra,Tongprl,longprl}.  For example, in 2001,  Knill
\emph{et al.} \cite{photon1} proposed a probabilistic scheme for
implementing a CNOT gate on two photonic qubits by using linear
optical elements, additional photons, and postselection. Based on
cross-Kerr nonlinearity or charge detection, Nemoto \emph{et al.}
\cite{photon5}, Lin \emph{et al.} \cite{photon9}, and Beenakker
\emph{et al.} \cite{MoveCNOT} provided some interesting proposals
for a deterministic quantum computing. In these schemes, some
additional qubits are employed. A strong cross-Kerr nonlinearity  is
still a big challenge in experiment at present. To achieve a
nontrivial nonlinearity between two individual qubits for a
deterministic quantum computation with the present experimental
techniques, an appealing platform for quantum information processing
with an artificial atom and a cavity is proposed \cite{Hu1,Hu2}.

A quantum system combining a cavity and an artificial atom, such as
a quantum dot (QD), a superconducting qubit, or a diamond
nitrogen-vacancy center, is a perfect platform for quantum
information processing because of its long coherence time and its
good  scalability. By  utilizing such a platform, some interesting
schemes were proposed for implementing the quantum gates on hybrid
photon-matter systems \cite{Hu1,Hu2,Hybrid1,Hybrid2}. Based on the
QD-cavity platform, a scalable deterministic quantum computation on
photonic qubits \cite{photoncomput1,photoncomput2,photoncomput3} and
a deterministic photonic spatial-polarization hyper-CNOT gate
\cite{HyperCNOT} were proposed recently. The quantum circuits for
the universal gates on superconducting qubits \cite{Super1,Super5}
or diamond nitrogen-vacancy center qubits \cite{NV1,NV2,NV3}
assisted by optical microcavities were  designed  as well.
Constructing universal quantum gates compactly can reduce the
quantum resource needed and their errors.

A QD system is one of the promising candidates for quantum
information processing  and quantum state storage in solid-state
quantum systems. The coherence time of a QD can be extended to
$\mu$s by using spin echo techniques
\cite{cohertime1,cohertime2,cohertime3}. The single QD spin
manipulation which is crucial for the implementation of single-qubit
gates, can be achieved by using pulsed magnetic resonance
techniques, nanosecond microwave pulses, or picosecond/femtosecond
optical pulses
\cite{spin-manipulate1,spin-manipulate2,spin-manipulate3}. Due to
the external magnetic field and the short dephasing time, the
magnetic resonance techniques are not compatible with our work. In
our work, the $90^\circ$  rotation on the electron-spin qubit around
the optical axis can be achieved by using a single photon, and the
$180^\circ$ rotation can be achieved by using  a single photon which
interacts with the QD twice \cite{Hu4}.

In this paper, we present some compact quantum circuits for a
universal quantum computing on an electron-spin system assisted by
the QDs inside single-side optical microcavities. Based on the giant
circular birefringence induced by a QD-cavity system as a result of
cavity quantum electrodynamics \cite{Hu1,Hu2}, we construct the
CNOT, Toffoli, and Fredkin gates on a stationary electron-spin
system, achieved  by some  input-output processes of a single
photon. Our schemes are simple and economic. They are accomplished
with a success probability of 100\% in principle and they do not
require the additional electron-spin qubits which are employed in
\cite{photon5,photon9,MoveCNOT}. Our circuits for implementing the
CNOT and Toffoli gates are especially compact. The electron qubits
involved in these gates are stationary, which reduces the
interaction between the spins and their environments, different from
\cite{MoveCNOT}. Moreover, our quantum circuits for the Toffoli and
Fredkin gates beat their synthesis with two-qubit entangling  gates
and single-qubit gates largely. With current technology, these
universal solid-state quantum gates are feasible.

\section{Compact quantum circuit for a CNOT gate on a stationary electron-spin system}
\label{sec2}

\subsection{A singly charged quantum dot in a single-side optical resonant microcavity} \label{sec2-1}

\begin{figure}[tpb]           
\begin{center}
\includegraphics[width=8 cm,angle=0]{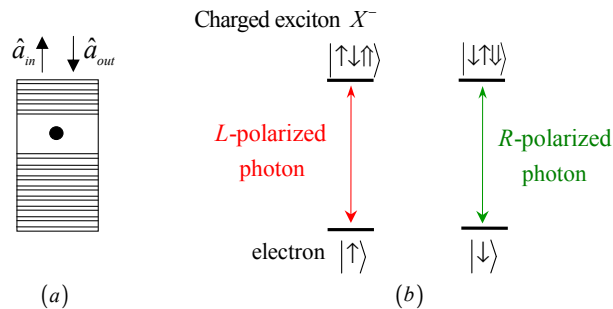}
\caption{ (a) Schematic diagram of a coupled single-side QD-cavity
system. (b) The energy-level structure of a QD-cavity system
\cite{Hu1,Hu2}.
$\vert\uparrow\rangle\rightarrow|\uparrow\downarrow\Uparrow\rangle$
is driven by the left-circularly polarized photon ($\vert L\rangle$)
and
$\vert\downarrow\rangle\rightarrow|\downarrow\uparrow\Downarrow\rangle$
is driven by the right-circularly polarized photon ($\vert
R\rangle$), respectively.} \label{level}
\end{center}
\end{figure}

Figure \ref{level} depicts the single-side QD-cavity system used in
our schemes, i.e., a self-assembled In(Ga)As QD or a GaAs interface
QD embedded in an optical resonant microcavity with one mirror
partially reflective and the another one  100\% reflective
\cite{Hu1,Hu2}. According to Pauli's exclusion principle, a
negatively charged exciton ($X^-$) consisting of two electrons bound
to one hole   can be optically excited when an excess electron is
injected into the QD \cite{HyperCNOT}.   In Fig. \ref{level},
$|\uparrow\rangle$ and $|\downarrow\rangle$ represent the spins of
the excess electron with the angular momentum projections $J_z=+1/2$
and $J_z=-1/2$ along the cavity axis, respectively.
$|\Uparrow\rangle$ and $|\Downarrow\rangle$ represent the hole-spin
states with $J_z=+3/2$ and $J_z=-3/2$, respectively. $|R\rangle$ and
$|L\rangle$ present the right-circularly polarized photon and the
left-circularly polarized photon, respectively.  In 2008,  Hu
\emph{et al.} \cite{Hu1,Hu2} showed that the L-polarized photon
($\vert L\rangle$) drives $|\uparrow\rangle$ transform into
$|\uparrow\downarrow\Uparrow\rangle$ and the R-polarized photon
($\vert R\rangle$) drives $|\downarrow\rangle$ transform into
$|\downarrow\uparrow\Downarrow\rangle$, respectively, due to Pauli's
exclusion principle.    The coupled $R$-polarized ($L$-polarized)
photon and the uncoupled  $L$-polarized ($R$-polarized) photon
acquire  different phases and amplitudes when they are  reflected by
the cavity. The reflection coefficient
\begin{eqnarray}       \label{eq1}
r(\omega)&=&|r(\omega)|e^{i\varphi(\omega)}=1-\frac{\kappa[i(\omega_{X^{-}}-\omega)+\frac{\gamma}{2}]}{[i(\omega_{X^{-}}-\omega)
+\frac{\gamma}{2}][i(\omega_c-\omega)+\frac{\kappa}{2}+\frac{\kappa_s}{2}]+g^2}
\end{eqnarray}
can be obtain by solving the Heisenberg equations of the motion for
the cavity mode $\hat{a}$ and the dipole operation $\sigma_{-}$
driven by the input field $\hat{a}_{in}$, and combing the relation
between the input field $\hat{a}_{in}$ and the output field
$\hat{a}_{out}$ in the weak excitation approximation
\cite{Heisenberg}
\begin{eqnarray}          \label{eq2}
\frac{d\hat{a}}{dt} &=&
-\left[i(\omega_c-\omega)+\frac{\kappa}{2}+\frac{\kappa_s}{2}\right]\hat{a}-g\sigma_{-}
-\sqrt{\kappa}\,\hat{a}_{in} +\hat{H}, \nonumber\\
\frac{d\sigma_-}{dt} &=& -\left[i(\omega_{X^-}-\omega)+\frac{\gamma}{2}\right]\sigma_{-}-g\sigma_z\,\hat{a}+\hat{G},\nonumber\\
\hat{a}_{\text{out}} &=& \hat{a}_{\text{in}}+\sqrt{\kappa}\,\hat{a}.
\end{eqnarray}
Here $\omega_c$ and $\omega$ are the frequencies of the cavity mode
and the input single photon, respectively. $\omega_{X^-}$ is the
frequency of the dipole transition of the negatively charged exciton
$X^-$.  $g$ is the coupling  strength between  the cavity mode and
$X^-$. $\kappa/2$ and $\kappa_s/2$ are the decay rate and the side
leakage rate of the cavity field,  respectively. $\gamma/2$
represents the decay rate of $X^-$.  $\hat{H}$ and $\hat{G}$ are the
noise operators related to the reservoirs.

Hu \emph{et al.} \cite{Hu1,Hu2} showed that
 $|r_0(\omega)|\simeq1$ for all $\omega$ if $\kappa_s\ll\kappa$.  If
$\kappa_s\ll\kappa$ and $g>(\kappa,\gamma)$,  one can see that
$|r_h(\omega)|\simeq1$ when $|\omega-\omega_c|\ll g$. Here
$r_0(\omega)$ and $r_h(\omega)$ are given by Eq. (\ref{eq1}) with
$g=0$ and $g\neq0$, respectively.  When $\kappa_s$ is negligible,
the transformations induced by the interaction between the QD and
the input single photon can be expressed as follows:
\begin{eqnarray}       \label{eq3}
(|R\rangle+|L\rangle)|\uparrow\rangle &\xrightarrow{\text{cav}}&
(e^{i\varphi_0}|R\rangle+e^{i\varphi_h}|L\rangle)|\uparrow\rangle=
e^{i\varphi_0}(|R\rangle+e^{i(\varphi_h-\varphi_0)}|L\rangle)|\uparrow\rangle,
\nonumber\\
(|R\rangle+|L\rangle)|\downarrow\rangle &\xrightarrow{\text{cav}}&
e^{i\varphi_h}|R\rangle|\downarrow\rangle
+e^{i\varphi_0}|L\rangle|\downarrow\rangle =
e^{i\varphi_0}(e^{i(\varphi_h-\varphi_0)}|R\rangle+|L\rangle)|\downarrow\rangle.\;\;\;\;\;\;
\end{eqnarray}
Here $\varphi_0=\arg[r_0(\omega)]$  and
$\varphi_h=\arg[r_h(\omega)]$. We consider the case that the QD is
resonant with the cavity mode and it interacts with the resonant
single photon (i.e., $\omega_{X^-}=\omega_c=\omega$) in the
conditions $\kappa_s\ll\kappa$ and $g>(\kappa,\gamma)$ below. In
this case, $e^{i\varphi_0}=-1$ and $e^{i\varphi_h}=1$. That is,
$\varphi_h-\varphi_0=\pm\pi$.  The rules of the input states
changing under the interaction of the photon and the cavity can be
summarized as follows:
\begin{eqnarray}       \label{eq4}
|R\rangle|\uparrow\rangle
&\xrightarrow{\text{cav}}&-|R\rangle|\uparrow\rangle,\;\;\;\;\;\;\;
|L\rangle|\uparrow\rangle\xrightarrow{\text{cav}}|L\rangle|\uparrow\rangle,\nonumber\\
|R\rangle|\downarrow\rangle&\xrightarrow{\text{cav}}&|R\rangle|\downarrow\rangle,\;\;\;\;\;\;\;\;\;\;
|L\rangle|\downarrow\rangle\xrightarrow{\text{cav}}-|L\rangle|\downarrow\rangle.\;\;\;\;\;\;\;
\end{eqnarray}
In 2011, Young \emph{et al.} \cite{Hu6} measured the macroscopic
phase shift of the reflected photon from a single-side pillar
microcavity induced by a single QD in experiment.  In a realistic
cavity system, although it is hard to achieve the  phase shift
$\varphi_h-\varphi_0=\pm\pi$ due to the side leakage and the cavity
loss \cite{hard}, the phase shift $\pm\pi/2$ can be actually
achieved in a QD-single-side-cavity system and it has been
demonstrated by Hu's group \cite{Hu4}. When $\kappa_s<1.3\kappa$,
the phase shift $\pm\pi/2$ can be achieved; otherwise, it cannot be
achieved.  The phase shift $\pi$ in our schemes can  be achieved by
a single photon which interacts with the QD twice. The above model
works for a general polarization-degenerate cavity mode, including
the micropillar
\cite{unpolarized-pillar1,unpolarized-pillar2,unpolarized-pillar3},
 H1 photonic crystal
\cite{unpolarized-photon1,unpolarized-photon2}, and fiber-based
\cite{fiber-based} cavities.

Utilizing the optical circular birefringence induced by cavity
quantum electrodynamics, the QD-cavity platform has been used to
generate the maximally entangled states
\cite{Hu1,Hu2,Hu3,Hu4,Hu5,Ren}, construct the conditional phase gate
on hybrid photon-QD systems \cite{Hu1,Hu2}, and  design the
hyper-CNOT gate on photonic qubits \cite{HyperCNOT}. Based on the
double-side one \cite{Hu3}, some universal quantum gates on photonic
qubits \cite{photoncomput1,photoncomput2} and hybrid photon-QD
systems \cite{Hybrid2} have been proposed. In 2011, Wang \emph{et
al.} \cite{repeater} proposed a scheme for implementing a quantum
repeater, resorting to the QDs in double-side cavities. In the
following, we discuss the implementation of a deterministic quantum
computing with QD-single-side-cavity systems, shown in Fig.
\ref{level}. The QD-double-side-cavity system is robust to the
transmission and the reflection coefficients, while the side leakage
rate of the QD-single-side-cavity system is lower than the
double-side one.

\begin{figure}[tpb]      
\begin{center}
\includegraphics[width=8 cm,angle=0]{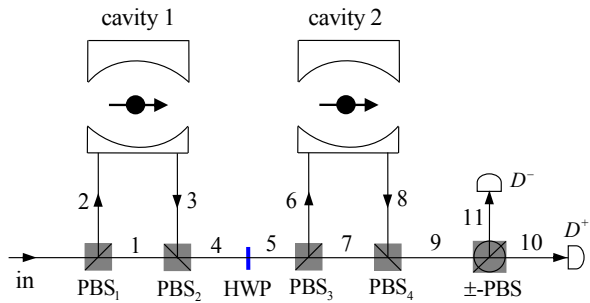}
\caption{Compact quantum circuit for deterministically implementing
a CNOT gate on two QD electron-spin qubits with  a single-photon
medium. The polarizing beam splitter PBS$_i$ ($i=1,2,3,4$) in the
basis $\{|R\rangle,\;|L\rangle\}$ transmits the $R$-polarized photon
and reflects the $L$-polarized photon. BS is a 50:50 beam splitter.
The $\pm$--PBS transmits the photon in the state
$|+\rangle=(|R\rangle+|L\rangle)/\sqrt{2}$  and reflects the photon
in the state $|-\rangle=(|R\rangle-|L\rangle)/\sqrt{2}$. The half
wave plate (HWP) set to 22.5$^\circ$ induces the transformations
$|R\rangle\xrightarrow{H_p}(|R\rangle+|L\rangle)/\sqrt{2}$ and
$|L\rangle\xrightarrow{H_p}(|R\rangle-|L\rangle)/\sqrt{2}$. $D^+$
and $D^-$ represent two single-photon detectors.} \label{CNOT}
\end{center}
\end{figure}

\subsection{Compact circuit for a  CNOT  gate on a stationary electron-spin system} \label{sec2-2}

The principle for implementing a CNOT gate on the two stationary
electron-spin qubits in the  QDs confined in single-side resonant
optical microcavities  is shown  in  Fig. \ref{CNOT}. It flips the
state of the target qubit when the control qubit is in the state
$|\downarrow\rangle$. Suppose the input state of the quantum system
composed of the control and the target qubits (confined in the
cavities 1 and 2, respectively) are initially prepared as
\begin{eqnarray}                    \label{eq5}
|\psi\rangle_{\text{in}}^{e} =
|\uparrow\rangle_{c}(\alpha_{1}|\uparrow\rangle_{t} +
\alpha_2|\downarrow\rangle_{t})
 + |\downarrow\rangle_{c}(\alpha_{3}|\uparrow\rangle_{t} + \alpha_4|\downarrow\rangle_{t}).
\end{eqnarray}
Here $\sum_{i=1}^4|\alpha_i|^2=1$. The input single photon  is
prepared in the equal polarization superposition state
$|\psi\rangle^{p} = \frac{1}{\sqrt{2}}(|R\rangle+|L\rangle)$.

Let us introduce the principle of our deterministic CNOT gate on two
stationary electron-spin qubits. As depicted in Fig. \ref{CNOT}, a
single photon  is injected into the input port $in$, and its
$R$-polarized component is transmitted to the spatial model 1 by the
polarizing beam splitter PBS$_1$ and then arrives at PBS$_2$
directly, while its $L$-polarized component is reflected to the
spatial model 2 for interacting with the QD inside the cavity 1.
After the photon emitting from the spatial models 1 and 3 arrives at
PBS$_2$ simultaneously, a Hadamard operation $H_p$ is performed on
it. That is, we let the photon pass through the half-wave plate
(HWP) oriented at 22.5$^\circ$, which results in the transformations
as follows:
\begin{eqnarray}                    \label{eq7}
|R\rangle \;\xrightarrow{H_p}\;
\frac{1}{\sqrt{2}}(|R\rangle+|L\rangle),\;\;\;\;\;\;\;\; |L\rangle
\;\xrightarrow{H_p}\; \frac{1}{\sqrt{2}}(|R\rangle-|L\rangle).
\end{eqnarray}
Before and after the photon passes through the block composed of
PBS$_3$, the QD inside the cavity 2, and PBS$_4$, a Hadamard
operation $H_e$ is performed on the electron spin in the QD inside
the cavity 2, respectively. Here $H_e$ completes the transformations
as follows:
\begin{eqnarray}                    \label{eq8}
|\uparrow\rangle \;\xrightarrow{H_e}\;
\frac{1}{\sqrt{2}}(|\uparrow\rangle+|\downarrow\rangle),\;\;\;\;\;\;\;\;
|\downarrow\rangle \;\xrightarrow{H_e}\;
\frac{1}{\sqrt{2}}(|\uparrow\rangle-|\downarrow\rangle).
\end{eqnarray}
The evolution of the whole system  composed of a single-photon
medium and the QDs inside the cavities 1 and 2 induced by the above
operations ($\text{PBS}_1$ $\rightarrow$ $\text{cavity 1}$
$\rightarrow$ $\text{PBS}_2$ $\rightarrow$ $\text{HWP}$
$\rightarrow$ $\text{H}_{e_2}$ $\rightarrow$ $\text{PBS}_3$
$\rightarrow$ $\text{cavity 2} \rightarrow \text{PBS}_4 \rightarrow
H_{e_2}$) can be
 described as follows:
\begin{eqnarray}                    \label{eq9}
|\psi\rangle^{p}\otimes|\psi\rangle_{\text{in}}^{e} & \rightarrow &
|R\rangle_9|\uparrow\rangle_{c}(\alpha_{1}|\uparrow\rangle_{t}+\alpha_2|\downarrow\rangle_{t})
+|L\rangle_9|\downarrow\rangle_{c}(\alpha_3|\downarrow\rangle_{t}+\alpha_4|\uparrow\rangle_{t}).
\end{eqnarray}
Here and below, we use $|R\rangle_i$ ($|L\rangle_i$) to denote the
photon in the state $|R\rangle$ ($|L\rangle$) emitting from the
spatial mode $i$ and use $H_{e_i}$ to denote a Hadamard operation
performed on the $i$-th QD-spin qubit.

Next, the single photon   is measured in the basis
$\{|\pm\rangle=(|R\rangle\pm|L\rangle)/\sqrt{2}\}$ by the detectors
$D^+$ and  $D^-$. From Eq. (\ref{eq9}), one can see that the
response of the detector $D^+$ indicates that the CNOT gate on the
two electron-spin qubits succeeds; if the detector $D^-$ is clicked,
after we perform a classical feed-forward operation
$\sigma_z=|\uparrow\rangle\langle\uparrow|-|\downarrow\rangle\langle\downarrow|$
on the control qubit,  the CNOT gate is accomplished as well.  That
is, the output state of the system composed of the control and the
target qubits confined in the cavities 1 and 2 becomes
\begin{eqnarray}                    \label{eq10}
|\psi\rangle_{\text{in}}^e \;\; \xrightarrow{\text{CNOT}} \;\;
|\psi\rangle_{\text{out}}^e  =
\alpha_{1}|\uparrow\rangle_{c}|\uparrow\rangle_{t}+\alpha_2|\uparrow\rangle_{c}|\downarrow\rangle_{t}
+\alpha_{3}|\downarrow\rangle_{c}|\downarrow\rangle_{t}+\alpha_4|\downarrow\rangle_{c}|\uparrow\rangle_{t}.\;\;\;\;\;\;
\end{eqnarray}
The quantum circuit shown in Fig. \ref{CNOT} can be used to
implement a CNOT gate on  the  two-qubit  electron-spin  system in a
deterministic way, which implements a not operation on the  target
qubit if and only if (iff) the control qubit is in the state
$|\downarrow\rangle$.

\section{Compact quantum circuit for  a Toffoli gate on three  electron-spin qubits in QDs} \label{sec3}

\begin{figure*}[tpb]   
\begin{center}
\includegraphics[width=13.2 cm,angle=0]{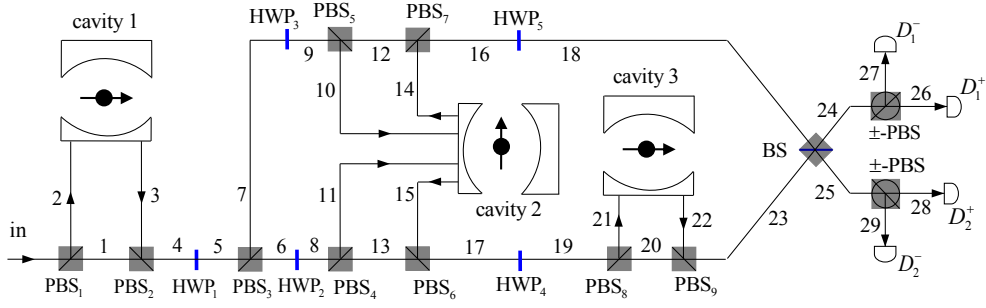}
\caption{Compact quantum circuit  for deterministically implementing
a Toffoli gate on three stationary electron-spin qubits in QDs with
the input-output processes of a single-photon medium.}
\label{Toffoli}
\end{center}
\end{figure*}

The principle for implementing a Toffoli gate on a three-qubit
electron-spin system is shown in Fig. \ref{Toffoli}. It is used to
flip the state of the target qubit iff both the two control qubits
are in the state $|\downarrow\rangle$. Suppose the quantum system,
which is composed of the three independent excess electrons inside
the cavities 1, 2, and 3 that act as the first control qubit, the
second control qubit, and the target qubit, respectively, is
initially prepared in an arbitrary state
\begin{eqnarray}                    \label{eq11}
\begin{split} |\Xi\rangle_{\text{in}}^e&=
|\uparrow\rangle_{c_1}|\uparrow\rangle_{c_2}(\alpha_{1}|\uparrow\rangle_t+\alpha_{2}|\downarrow\rangle_t)
+|\uparrow\rangle_{c_1}|\downarrow\rangle_{c_2}(\alpha_{3}|\uparrow\rangle_t+\alpha_{4}|\downarrow\rangle_t)
\\&\;\;\;\;
+|\downarrow\rangle_{c_1}|\uparrow\rangle_{c_2}(\alpha_{5}|\uparrow\rangle_t+\alpha_{6}|\downarrow\rangle_t)
+|\downarrow\rangle_{c_1}|\downarrow\rangle_{c_2}(\alpha_{7}|\uparrow\rangle_t+\alpha_{8}|\downarrow\rangle_t).
\end{split}
\end{eqnarray}
Here $\sum_{i=1}^8|\alpha_{i}|^2=1$.

Next, we will specify the evolution of the system from the input
state to the output state for characterizing the performance of our
Toffoli gate. As illustrated in Fig. \ref{Toffoli}, our scheme for a
Toffoli gate on a three-qubit electron-spin system  can be achieved
with four steps.

First, an input single photon  in the  state  $|\Xi\rangle^{p} =
\frac{1}{\sqrt{2}}(|R\rangle-|L\rangle)$ goes through the block
composed of PBS$_1$, the QD inside the cavity 1, and PBS$_2$, and
then  an $H_p$ is performed on it (i.e., let the photon go though
HWP$_1$). Based on the argument as  made in Sec. \ref{sec2-2}, one
can see that the above operations
($\text{PBS}_1\rightarrow\text{cavity\;1}\rightarrow\text{PBS}_2\rightarrow\text{HWP}_1$)
transform the state of the complicated  system  composed of the
single photon and the three QD-spin qubits from $|\Xi_0\rangle$ into
$|\Xi_1\rangle$. Here
\begin{eqnarray}                      \label{eq13}
|\Xi_0\rangle&=&|\Xi\rangle^{p}\otimes|\Xi\rangle_{\text{in}}^e,\nonumber\\
|\Xi_1\rangle&=&
|L\rangle_5|\uparrow\rangle_{c_1}|\uparrow\rangle_{c_2}(\alpha_{1}|\uparrow\rangle_{t}+\alpha_{2}|\downarrow\rangle_{t})
+|L\rangle_5|\uparrow\rangle_{c_1}|\downarrow\rangle_{c_2}(\alpha_{3}|\uparrow\rangle_{t}+\alpha_{4}|\downarrow\rangle_{t})\nonumber\\&&
+|R\rangle_5|\downarrow\rangle_{c_1}|\uparrow\rangle_{c_2}(\alpha_{5}|\uparrow\rangle_{t}+\alpha_{6}|\downarrow\rangle_{t})
+|R\rangle_5|\downarrow\rangle_{c_1}|\downarrow\rangle_{c_2}(\alpha_{7}|\uparrow\rangle_{t}+\alpha_{8}|\downarrow\rangle_{t}).
\end{eqnarray}

Second, PBS$_3$ transforms $|R\rangle_5$ and  $|L\rangle_5$ into
$|R\rangle_6$  and   $|L\rangle_7$, respectively.  Before and after
the component $|R\rangle_6$ ($|L\rangle_7$) of the photon goes
through the block composed of PBS$_4$, the QD inside the cavity 2,
and PBS$_6$ (PBS$_5$, the QD inside the cavity 2, and PBS$_7$), an
$H_p$ is performed on it  with HWP$_2$ and HWP$_4$ (HWP$_3$ and
HWP$_5$). The operations
($\text{HWP}_2\rightarrow\text{PBS}_4\rightarrow\text{cavity\;2}\rightarrow\text{PBS}_6\rightarrow\text{HWP}_4$
and
$\text{HWP}_3\rightarrow\text{PBS}_5\rightarrow\text{cavity\;2}\rightarrow\text{PBS}_7\rightarrow\text{HWP}_5$)
transform the state of the  complicated system into
\begin{eqnarray}                      \label{eq14}
\rightarrow|\Xi_2\rangle&=&
|L\rangle_{18}|\uparrow\rangle_{c_1}|\uparrow\rangle_{c_2}(\alpha_{1}|\uparrow\rangle_{t}+\alpha_{2}|\downarrow\rangle_{t})
+|R\rangle_{18}|\uparrow\rangle_{c_1}|\downarrow\rangle_{c_2}(\alpha_{3}|\uparrow\rangle_{t}+\alpha_{4}|\downarrow\rangle_{t})
\nonumber\\&&
+|R\rangle_{19}|\downarrow\rangle_{c_1}|\uparrow\rangle_{c_2}(\alpha_{5}|\uparrow\rangle_{t}+\alpha_{6}|\downarrow\rangle_{t})
+|L\rangle_{19}|\downarrow\rangle_{c_1}|\downarrow\rangle_{c_2}(\alpha_{7}|\uparrow\rangle_{t}+\alpha_{8}|\downarrow\rangle_{t}).
\end{eqnarray}

Third, before and after the photon goes through the block composed
of PBS$_8$, the QD inside the cavity 3, and PBS$_9$  when it emits
from the spatial model 19, an $H_e$ is performed on the electron
spin in the QD inside the cavity 3, respectively. These operations
($H_{e_3}\rightarrow\text{PBS}_8\rightarrow\text{cavity\;3}\rightarrow\text{PBS}_9\rightarrow
H_{e_3}$) complete the transformation
\begin{eqnarray}                      \label{eq15}
\rightarrow|\Xi_3\rangle&=&
|L\rangle_{18}|\uparrow\rangle_{c_1}|\uparrow\rangle_{c_2}(\alpha_{1}|\uparrow\rangle_{t}+\alpha_{2}|\downarrow\rangle_{t})
+|R\rangle_{18}|\uparrow\rangle_{c_1}|\downarrow\rangle_{c_2}(\alpha_{3}|\uparrow\rangle_{t}+\alpha_{4}|\downarrow\rangle_{t})
\nonumber\\&&
+|R\rangle_{23}|\downarrow\rangle_{c_1}|\uparrow\rangle_{c_2}(\alpha_{5}|\uparrow\rangle_{t}+\alpha_{6}|\downarrow\rangle_{t})
+|L\rangle_{23}|\downarrow\rangle_{c_1}|\downarrow\rangle_{c_2}(\alpha_{7}|\downarrow\rangle_{t}+\alpha_{8}|\uparrow\rangle_{t}).
\end{eqnarray}
Subsequently, the wave packet emitting from the spatial model 23
arrives at the 50:50 beam splitter (BS) with the wave packet
emitting from the spatial model 18 simultaneously.

Fourth, the balanced BS, which completes the transformations
\begin{eqnarray}                      \label{eq16}
|R\rangle_{18}&\xrightarrow{\text{BS}}&\frac{1}{\sqrt{2}}(|R\rangle_{24}+|R\rangle_{25}),\;\;\;\;\;\;\;\;\;
|L\rangle_{18}\;\xrightarrow{\text{BS}}\;\;\frac{1}{\sqrt{2}}(|L\rangle_{24}+|L\rangle_{25}),\nonumber\\
|R\rangle_{23}&\xrightarrow{\text{BS}}&\frac{1}{\sqrt{2}}(|R\rangle_{24}-|R\rangle_{25}),\;\;\;\;\;\;\;\;\;
|L\rangle_{23}\;\;\xrightarrow{\text{BS}}\;\;\frac{1}{\sqrt{2}}(|L\rangle_{24}-|L\rangle_{25}),
\end{eqnarray}
transforms $|\Xi_3\rangle$ into the state
\begin{eqnarray}                      \label{eq17}
\xrightarrow{\text{BS}}|\Xi_4\rangle & = & \frac{|+\rangle_{26}}{2}
\Big[|\uparrow\rangle_{c_1}|\uparrow\rangle_{c_2}(\alpha_{1}|\uparrow\rangle_t+\alpha_{2}|\downarrow\rangle_t)
     +|\uparrow\rangle_{c_1}|\downarrow\rangle_{c_2}(\alpha_{3}|\uparrow\rangle_t+\alpha_{4}|\downarrow\rangle_t)\nonumber\\&
     & +|\downarrow\rangle_{c_1}|\uparrow\rangle_{c_2}(\alpha_5|\uparrow\rangle_t+\alpha_{6}|\downarrow\rangle_t)
     +|\downarrow\rangle_{c_1}|\downarrow\rangle_{c_2}(\alpha_{7}|\downarrow\rangle_t+\alpha_{8}|\uparrow\rangle_t)\Big]\nonumber\\&
&+\frac{|-\rangle_{27}}{2}
\Big[-|\uparrow\rangle_{c_1}|\uparrow\rangle_{c_2}(\alpha_{1}|\uparrow\rangle_t+\alpha_{2}|\downarrow\rangle_t)
    +|\uparrow\rangle_{c_1}|\downarrow\rangle_{c_2}(\alpha_{3}|\uparrow\rangle_t+\alpha_{4}|\downarrow\rangle_t)\nonumber\\&&
    +|\downarrow\rangle_{c_1}|\uparrow\rangle_{c_2}(\alpha_{5}|\uparrow\rangle_t+\alpha_{6}|\downarrow\rangle_t)
    -|\downarrow\rangle_{c_1}|\downarrow\rangle_{c_2}(\alpha_{7}|\downarrow\rangle_t+\alpha_{8}|\uparrow\rangle_t)\Big]\nonumber\\&&
    +\frac{|+\rangle_{28}}{2}
\Big[|\uparrow\rangle_{c_1}|\uparrow\rangle_{c_2}(\alpha_{1}|\uparrow\rangle_t+\alpha_{2}|\downarrow\rangle_t)
   +|\uparrow\rangle_{c_1}|\downarrow\rangle_{c_2}(\alpha_{3}|\uparrow\rangle_t+\alpha_{4}|\downarrow\rangle_t)\nonumber\\&&
    -|\downarrow\rangle_{c_1}|\uparrow\rangle_{c_2}(\alpha_{5}|\uparrow\rangle_t+\alpha_{6}|\downarrow\rangle_t)
   -|\downarrow\rangle_{c_1}|\downarrow\rangle_{c_2}(\alpha_{7}|\downarrow\rangle_t+\alpha_{8}|\uparrow\rangle_t)\Big]\nonumber\\&&
   +\frac{|-\rangle_{29}}{2}
\Big[-|\uparrow\rangle_{c_1}|\uparrow\rangle_{c_2}(\alpha_{1}|\uparrow\rangle_t+\alpha_{2}|\downarrow\rangle_t)
   +|\uparrow\rangle_{c_1}|\downarrow\rangle_{c_2}(\alpha_{3}|\uparrow\rangle_t+\alpha_{4}|\downarrow\rangle_t)\nonumber\\&&
   -|\downarrow\rangle_{c_1}|\uparrow\rangle_{c_2}(\alpha_{5}|\uparrow\rangle_t+\alpha_{6}|\downarrow\rangle_t)
   +|\downarrow\rangle_{c_1}|\downarrow\rangle_{c_2}(\alpha_{7}|\downarrow\rangle_t+\alpha_{8}|\uparrow\rangle_t)\Big].
\end{eqnarray}
According to the outcomes of the measurement on the single photon
 in the basis $\{|\pm\rangle\}$, we perform the appropriate
single-qubit operations on the qubits shown in Table
\ref{tableToffoli}, and then the state of the solid-state quantum
system composed of the three electron-spin qubits becomes
\begin{eqnarray}                      \label{eq18}
|\Xi\rangle_{\text{out}}^e&=&
|\uparrow\rangle_{c_1}|\uparrow\rangle_{c_2}(\alpha_1|\uparrow\rangle_t+\alpha_2|\downarrow\rangle_t)
+|\uparrow\rangle_{c_1}|\downarrow\rangle_{c_2}(\alpha_3|\uparrow\rangle_t+\alpha_4|\downarrow\rangle_t)\nonumber\\
&&+|\downarrow\rangle_{c_1}|\uparrow\rangle_{c_2}(\alpha_5|\uparrow\rangle_t+\alpha_6|\downarrow\rangle_t)
+|\downarrow\rangle_{c_1}|\downarrow\rangle_{c_2}(\alpha_7|\downarrow\rangle_t+\alpha_8|\uparrow\rangle_t).
\end{eqnarray}

From Eqs. (\ref{eq13}) and (\ref{eq18}), one can see that the
evolution
$|\Xi\rangle_{\text{in}}^e\xrightarrow{\text{Toffoli}}|\Xi\rangle_{\text{out}}^e$
is accomplished. That is, the quantum circuit  shown in Fig.
\ref{Toffoli} implements a Toffoli gate on the three stationary
electron-spin qubits in QDs, and it flips the state of the target
qubit inside the cavity 3 iff both the two control qubits inside the
cavities 1 and 2, respectively, are in the state
$|\downarrow\rangle$ with a successful probability of 100\% in
principle.

\begin{table}[htb]
\centering \caption{The relations between the measurement outcomes
of   the single photon and the classical feed-forward operations for
implementing the Toffoli gate on the three stationary electron-spin
qubits.
$\sigma_z=|\uparrow\rangle\langle\uparrow|-|\downarrow\rangle\langle\downarrow|$.
$I_2=|\uparrow\rangle\langle\uparrow|+|\downarrow\rangle\langle\downarrow|$
is a 2 $\times$ 2 unit operation which means doing nothing on a
qubit.}
\begin{tabular}{cccc}
\hline  \hline

           & \multicolumn {3}{c}{Feed-forward} \\
\cline{2-4}
$\;\;$photon  $\;\;$               & $\;\;$ qubit $c_1$       & $\;\;$ $\;\;$ qubit  $c_2$ $\;\;$   &$\;$  qubit $t$ $\;\;$\\
\hline
$D_1^+ (|+\rangle_{26})$            &  $I_2$                  & $I_2$                               &  $I_2$   \\
$D_1^-(|-\rangle_{27})$             &  $-\sigma_z$            &  $\sigma_z$                          &  $I_2$   \\
$D_2^+(|+\rangle_{28})$             &  $\sigma_z$              &  $I_2$                              &  $I_2$   \\
$D_2^-(|-\rangle_{29})$             &  $I_2$                  &  $-\sigma_z$                        &  $I_2$   \\
                             \hline  \hline
\end{tabular}\label{tableToffoli}
\end{table}

\section{Compact quantum circuit for a Fredkin gate on a three-qubit electron-spin system} \label{sec4}

\begin{figure*}[tpb]           
\begin{center}
\includegraphics[width=13 cm,angle=0]{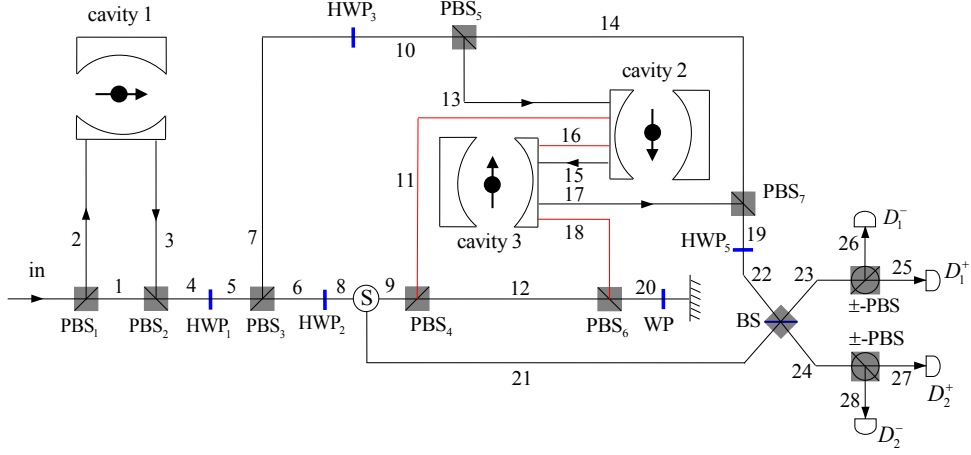}
\caption{Compact quantum circuit  for determinately implementing a
Fredkin gate on three QD-spin qubits with the input-output processes
of a single-photon medium. Wave plate WP performs a Hadamard
operation on the photon who goes through it two times in
succession.} \label{Fredkin}
\end{center}
\end{figure*}

Figure \ref{Fredkin} depicts the principle of our scheme for
implementing a Fredkin gate on  a three-qubit electron-spin system
assisted by the QDs inside single-side optical microcavities,  which
swaps the states of the two target qubits iff the control qubit is
in the state $|\downarrow\rangle$. Suppose the input state of the
system composed of the control qubit, the first target qubit, and
the second target qubit inside the cavities 1, 2, and 3,
respectively, is initially prepared as
\begin{eqnarray}                      \label{eq20}
|\Pi\rangle^e_{\text{in}}&=&
|\uparrow\rangle_{c}|\uparrow\rangle_{t_1}(\alpha_{1}|\uparrow\rangle_{t_2}+\alpha_{2}|\downarrow\rangle_{t_2})
+|\uparrow\rangle_{c}|\downarrow\rangle_{t_1}(\alpha_{3}|\uparrow\rangle_{t_2}+\alpha_{4}|\downarrow\rangle_{t_2})
\nonumber\\&&
+|\downarrow\rangle_{c}|\uparrow\rangle_{t_1}(\alpha_{5}|\uparrow\rangle_{t_2}+\alpha_{6}|\downarrow\rangle_{t_2})
+|\downarrow\rangle_{c}|\downarrow\rangle_{t_1}(\alpha_{7}|\uparrow\rangle_{t_2}+\alpha_{8}|\downarrow\rangle_{t_2}).
\end{eqnarray}
Here $\sum_{i=1}^8|\alpha_i|^2=1$. The input single photon  is
prepared in the state
$|\Pi\rangle^p=\frac{1}{\sqrt{2}}(|R\rangle-|L\rangle)$.

Let us now describe the principle of our scheme for implementing a
Fredkin gate on the three stationary electron-spin qubits in QDs  in
detail.

First, based on the argument as  made in Sec. \ref{sec3}, after the
input photon   goes through the block composed of PBS$_1$, the QD
inside the cavity 1, and PBS$_2$, an $H_p$ (i.e., let it go through
HWP$_1$) is performed on it, and then the state of the whole system
composed of the single photon and the three electron-spin qubits in
the QDs confined in the cavities 1, 2, and 3 is transformed from
$|\Pi_0\rangle$ into $|\Pi_1\rangle$ by the above operations
($\text{PBS}_1\rightarrow\text{cavity\;1}\rightarrow\text{PBS}_2\rightarrow\text{HWP}_1$).
Here
\begin{eqnarray}                      \label{eq21}
|\Pi_0\rangle&=& |\Pi\rangle^p\otimes|\Pi\rangle^e_{\text{in}}, \nonumber\\
|\Pi_1\rangle&=&
|L\rangle_5|\uparrow\rangle_{c}|\uparrow\rangle_{t_1}(\alpha_{1}|\uparrow\rangle_{t_2}+\alpha_{2}|\downarrow\rangle_{t_2})
+|L\rangle_5|\uparrow\rangle_{c}|\downarrow\rangle_{t_1}(\alpha_{3}|\uparrow\rangle_{t_2}+\alpha_{4}|\downarrow\rangle_{t_2})\nonumber\\&&
+|R\rangle_5|\downarrow\rangle_{c}|\uparrow\rangle_{t_1}(\alpha_{5}|\uparrow\rangle_{t_2}+\alpha_{6}|\downarrow\rangle_{t_2})
+|R\rangle_5|\downarrow\rangle_{c}|\downarrow\rangle_{t_1}(\alpha_{7}|\uparrow\rangle_{t_2}+\alpha_{8}|\downarrow\rangle_{t_2}).
\end{eqnarray}

Second, PBS$_3$ transforms $|R\rangle_5$ and  $|L\rangle_5$  into
$|R\rangle_6$  and  $|L\rangle_7$, respectively. When the photon is
in the state $|L\rangle_7$, before and after it goes through the
block composed of PBS$_5$, the QDs inside the cavities 2 and 3, and
PBS$_7$, an $H_p$ is performed on it with HWP$_3$ and HWP$_5$,
respectively, and then it arrives at the balanced BS directly. When
the photon is in the state $|R\rangle_6$, after an $H_p$ is
performed on it with HWP$_2$, the optical switch $S$ leads it to the
block composed of PBS$_4$, the QDs inside the cavities 2 and 3, and
PBS$_6$, following with an $H_p$ which is performed on the photon
with a wave plate (WP) and a mirror. Here
$|R\rangle_{20}\xrightarrow{\text{WP}}\xrightarrow{\text{mirror}}\xrightarrow{\text{WP}}(|R\rangle_{20}+|L\rangle_{20})/\sqrt{2}$
and $|L\rangle_{20}\xrightarrow{\text{WP}}\xrightarrow{  \text{
mirror}}\xrightarrow{\text{WP}}(|R\rangle_{20}-|L\rangle_{20})/\sqrt{2}$.
These operations
($\text{HWP}_3\rightarrow\text{PBS}_5\rightarrow\text{cavity\;2}\rightarrow\text{cavity\;3}\rightarrow\text{PBS}_7\rightarrow\text{HWP}_5$
and $\text{HWP}_2\rightarrow
S\rightarrow\text{PBS}_4\rightarrow\text{cavity\;2}\rightarrow\text{cavity\;3}
\rightarrow\text{PBS}_6\rightarrow\text{WP}\rightarrow\text{mirror}\rightarrow\text{WP}$)
complete the transformation
\begin{eqnarray}                      \label{eq23}
\rightarrow|\Xi_2\rangle&=&
|\uparrow\rangle_{c}|\uparrow\rangle_{t_1}(\alpha_{1}|L\rangle_{22}|\uparrow\rangle_{t_2}+\alpha_{2}|R\rangle_{22}|\downarrow\rangle_{t_2})
+|\uparrow\rangle_{c}|\downarrow\rangle_{t_1}(\alpha_{1}|R\rangle_{22}|\uparrow\rangle_{t_2}
+\alpha_{2}|L\rangle_{22}|\downarrow\rangle_{t_2})\nonumber\\&&
+|\downarrow\rangle_{c}|\uparrow\rangle_{t_1}(\alpha_{1}|R\rangle_{20}|\uparrow\rangle_{t_2}
+\alpha_{2}|L\rangle_{20}|\downarrow\rangle_{t_2})
+|\downarrow\rangle_{c}|\downarrow\rangle_{t_1}(\alpha_{1}|L\rangle_{20}|\uparrow\rangle_{t_2}
+\alpha_{2}|R\rangle_{20}|\downarrow\rangle_{t_2}).
\end{eqnarray}

Third, the photon emitting from the spatial model 20 is injected
into the block composed of PBS$_6$, the QDs inside the cavities 2
and 3, and PBS$_4$ again, and before and after the photon interacts
with the QDs inside the cavities 3 and 2, an $H_e$ is performed on
the QDs inside the cavities 3 and 2, respectively. The optical
switch $S$ leads the wave packet to the spatial model 21 for
interfering with the wave packet emitting from the spatial model 22.
The above operations
($H_{e_2},H_{e_3}\rightarrow\text{PBS}_6\rightarrow
\text{cavity\;3}\rightarrow\text{cavity\;2}\rightarrow\text{PBS}_4\rightarrow
H_{e_2},H_{e_3}\rightarrow S$) complete
 the transformation
\begin{eqnarray}                      \label{eq24}
\rightarrow|\Xi_3\rangle&=&
|\uparrow\rangle_{c}|\uparrow\rangle_{t_1}(\alpha_{1}|L\rangle_{22}|\uparrow\rangle_{t_2}+\alpha_{2}|R\rangle_{22}|\downarrow\rangle_{t_2})
+|\uparrow\rangle_{c}|\downarrow\rangle_{t_1}(\alpha_{3}|R\rangle_{22}|\uparrow\rangle_{t_2}
+\alpha_{4}|L\rangle_{22}|\downarrow\rangle_{t_2})\nonumber\\&&
+|\downarrow\rangle_{c}(\alpha_{5}|R\rangle_{21}|\uparrow\rangle_{t_1}
+\alpha_{6}|L\rangle_{21}|\downarrow\rangle_{t_1})|\uparrow\rangle_{t_2}
+|\downarrow\rangle_{c}(\alpha_{7}|L\rangle_{21}|\uparrow\rangle_{t_1}
+\alpha_{8}|R\rangle_{21}|\downarrow\rangle_{t_1})|\downarrow\rangle_{t_2}.
\end{eqnarray}

Fourth, the single photon  is detected by the detectors $D_i^\pm$ in
the basis $\{|\pm\rangle\}$ after the 50:50 BS transforms
$|\Xi_3\rangle$ into $|\Xi_4\rangle$. Here
\begin{eqnarray}                      \label{eq25}
|\Xi_4\rangle &= \frac{|+\rangle_{25}}{2}
\Big[|\uparrow\rangle_{c}|\uparrow\rangle_{t_1}(\alpha_{1}|\uparrow\rangle_{t_2}+\alpha_{2}|\downarrow\rangle_{t_2})
     +|\uparrow\rangle_{c}|\downarrow\rangle_{t_1}(\alpha_{3}|\uparrow\rangle_{t_2}+\alpha_{4}|\downarrow\rangle_{t_2}) \nonumber\\&
     \;\;\;\;+|\downarrow\rangle_{c}(\alpha_5|\uparrow\rangle_{t_1}+\alpha_{6}|\downarrow\rangle_{t_1})|\uparrow\rangle_{t_2}
     +|\downarrow\rangle_{c}(\alpha_{7}|\uparrow\rangle_{t_1} +\alpha_{8}|\downarrow\rangle_{t_1})|\downarrow\rangle_{t_2}\Big] \nonumber\\&\;\;\;\;
+\frac{|-\rangle_{26}}{2}
\Big[|\uparrow\rangle_{c}|\uparrow\rangle_{t_1}(-\alpha_{1}|\uparrow\rangle_{t_2}+\alpha_{2}|\downarrow\rangle_{t_2})
     +|\uparrow\rangle_{c}|\downarrow\rangle_{t_1}(\alpha_{3}|\uparrow\rangle_{t_2} -\alpha_{4}|\downarrow\rangle_{t_2}) \nonumber\\&\;\;\;\;
     +|\downarrow\rangle_{c}(\alpha_5|\uparrow\rangle_{t_1}-\alpha_{6}|\downarrow\rangle_{t_1})|\uparrow\rangle_{t_2}
     +|\downarrow\rangle_{c}(-\alpha_{7}|\uparrow\rangle_{t_1}+\alpha_{8}|\downarrow\rangle_{t_1})|\downarrow\rangle_{t_2}\Big] \nonumber\\&\;\;\;\;
+\frac{|+\rangle_{27}}{2}
\Big[|\uparrow\rangle_{c}|\uparrow\rangle_{t_1}(\alpha_{1}|\uparrow\rangle_{t_2}+\alpha_{2}|\downarrow\rangle_{t_2})
     +|\uparrow\rangle_{c}|\downarrow\rangle_{t_1}(\alpha_{3}|\uparrow\rangle_{t_2}+\alpha_{4}|\downarrow\rangle_{t_2})\nonumber\\&\;\;\;\;
     -|\downarrow\rangle_{c}(\alpha_5|\uparrow\rangle_{t_1}+\alpha_{6}|\downarrow\rangle_{t_1})|\uparrow\rangle_{t_2}
     -|\downarrow\rangle_{c}(\alpha_{7}|\uparrow\rangle_{t_1}+\alpha_{8}|\downarrow\rangle_{t_1})|\downarrow\rangle_{t_2}\Big] \nonumber\\&\;\;\;\;
+\frac{|-\rangle_{28}}{2}
\Big[|\uparrow\rangle_{c}|\uparrow\rangle_{t_1}(-\alpha_{1}|\uparrow\rangle_{t_2}+\alpha_{2}|\downarrow\rangle_{t_2})
     +|\uparrow\rangle_{c}|\downarrow\rangle_{t_1}(\alpha_{3}|\uparrow\rangle_{t_2}-\alpha_{4}|\downarrow\rangle_{t_2}) \nonumber\\&\;\;\;\;
     +|\downarrow\rangle_{c}(-\alpha_5|\uparrow\rangle_{t_1} +\alpha_{6}|\downarrow\rangle_{t_1})|\uparrow\rangle_{t_2}
     +|\downarrow\rangle_{c}(\alpha_{7}|\uparrow\rangle_{t_1} -\alpha_{8}|\downarrow\rangle_{t_1})|\downarrow\rangle_{t_2}\Big]. 
\end{eqnarray}

Fifth, according to the outcomes of the measurement on the output
single photon, we perform  some appropriate classical  feed-forward
single-qubit operations, shown in Table \ref{tableFredkin},
 on the electron-spin qubits to make the state of the
system composed of the three electrons inside the cavities 1, 2, and
3 to be
\begin{eqnarray}                      \label{eq26}
|\Pi\rangle_{\text{out}}^e&=
|\uparrow\rangle_{c}(\alpha_{1}|\uparrow\rangle_{t_1}|\uparrow\rangle_{t_2}+\alpha_{2}|\uparrow\rangle_{t_1}|\downarrow\rangle_{t_2})
     +|\uparrow\rangle_{c}(\alpha_{3}|\downarrow\rangle_{t_1}|\uparrow\rangle_{t_2}+\alpha_{4}|\downarrow\rangle_{t_1}|\downarrow\rangle_{t_2})
     \nonumber\\&\;\;\;\;
     +|\downarrow\rangle_{c}(\alpha_5|\uparrow\rangle_{t_1}|\uparrow\rangle_{t_2}+\alpha_{6}|\downarrow\rangle_{t_1}|\uparrow\rangle_{t_2})
     +|\downarrow\rangle_{c}(\alpha_{7}|\uparrow\rangle_{t_1}|\downarrow\rangle_{t_2}
     +\alpha_{8}|\downarrow\rangle_{t_1}|\downarrow\rangle_{t_2}).
\end{eqnarray}

From Eqs. (\ref{eq21}) and (\ref{eq26}), one can see that the
evolution
$|\Pi\rangle_{\text{in}}\xrightarrow{\text{Fredkin}}|\Pi\rangle_{\text{out}}$
is completed. That is, the quantum circuit shown in Fig.
\ref{Fredkin} implements a Fredkin gate on the three-qubit
electron-spin system  in a deterministic way, which swaps the states
of the two target qubits iff the state of the control qubit is in
the state $|\downarrow\rangle$.

\begin{table}[htb]
\centering \caption{The relations between the measurement  outcomes
of  the photon  and the feed-forward operations for achieving a
Fredkin gate on the three-qubit electron-spin system. }
\begin{tabular}{cccc}
\hline  \hline

           & \multicolumn {3}{c}{Feed-forward} \\
\cline{2-4}
photon                        & $\;\;$ qubit $c$ $\;\;$     &  $\;\;$ qubit $t_1$    $\;\;\;$   &   $\;\;$ qubit $t_2$  $\;\;$ \\
\hline
$D_1^+(|+\rangle_{25})$  &   $I_2$             &   $I_2$               &    $I_2$           \\
$D_1^-(|-\rangle_{26})$  &   $-\sigma_z$       &   $\sigma_z$          &    $\sigma_z$                \\
$D_2^+(|+\rangle_{27})$  &   $\sigma_z$        &   $I_2$               &    $I_2$          \\
$D_2^-(|-\rangle_{28})$  &   $I_2$             &   $\sigma_z$          &    $\sigma_z$               \\

                             \hline  \hline
\end{tabular}\label{tableFredkin}
\end{table}

\section{The feasibilities and efficiencies of our schemes}\label{sec5}

So far, all the procedures in our schemes for the three universal
quantum gates are described  in  the  case that  the  side  leakage
rate $k_s$ is negligible. To present our ideas more realistically,
$k_s$ should be taken into account. In this time, the rules of the
input states changing under the interaction of the photon and the
cavity  become
\begin{eqnarray}       \label{eq27}
|R\rangle|\uparrow\rangle &\xrightarrow{\text{cav}}&
-|r_0||R\rangle|\uparrow\rangle,\;\;\;\;\;\;\;\;\;\;\;\;\;\;\;\;\;
|L\rangle|\uparrow\rangle\;\xrightarrow{\text{cav}}\;|r_h||L\rangle|\uparrow\rangle, \nonumber\\
|R\rangle|\downarrow\rangle &\xrightarrow{\text{cav}}&
|r_h||R\rangle|\downarrow\rangle,\;\;\;\;\;\;\;\;\;\;\;\;\;\;\;\,\;\;\;\;
|L\rangle|\downarrow\rangle\;\xrightarrow{\text{cav}}\;-|r_0||L\rangle|\downarrow\rangle.
\end{eqnarray}
The fidelities and the efficiencies of the universal quantum gates
are  sensitive to $k_s$ as $k_s$ influences the amplitudes of the
reflected photon (see Eq. (\ref{eq1})). Here the fidelity of a
quantum gate is defined as
\begin{eqnarray}                      \label{eq28}
F=|\langle\Psi_{\text{real}}|\Psi_{\text{ideal}}\rangle|^2,
\end{eqnarray}
where  $|\Psi_{\text{ideal}}\rangle$ is the output state of the
system composed of the QD-spin qubits involved in the gate and a
single-photon medium in the ideal case (that is,  the photon escapes
through the input-output mode). $|\Psi_{\text{real}}\rangle$ is the
output state of the complicated system in the realistic case (that
is, the cavities are imperfect and the side leakage $\kappa_s$ is
taken into account). The efficiency of the gate is considered as
\begin{eqnarray}                      \label{eq29}
\eta=n_{\text{out}}/n_{\text{in}}.
\end{eqnarray}
Here $n_{\text{in}}$ and $n_{\text{out}}$ are the numbers of the
input photons and the output photons, respectively.

For perfect cavities, the fidelities of our universal quantum gates
can reach  unity. By considering the side leakage and combining the
specific processes of the construction for the universal quantum
gates discussed above, the fidelities of our CNOT gate $F_{C}$,
Toffoli gate $F_{T}$, and Fredkin gate $F_{F}$, and their
efficiencies $\eta_{C}$, $\eta_{T}$, and $\eta_{F}$ can be
calculated as follows:
\begin{eqnarray}                      \label{eq30}
F_{C}=\frac{1}{2}\times\big(1\!+\!2|r_h|\!+\!|r_0||r_h|\big)/\big[(1\!+\!|r_h|)^2\!+\!(1\!-\!|r_0|)^2\!+\!|r_h|^2(1\!-\!|r_h|)^2\!+\!|r_h|^2(1\!+\!|r_0|)^2\big],
\end{eqnarray}
\begin{eqnarray}                      \label{eq31}
F_{T}& =\frac{1}{4}\times \big(3 + 2 |r_0| + |r_h| \big[5 + |r_h| +
|r_0|(4+|r_0|)\big]\big)/ \big((1 + |r_h|)^4 \nonumber\\&\;\;\;\; +
2 (|r_h|^2-1)^2+ 2 (|r_h|-1)^2(|r_0|-1)^2 + (|r_0|-1)^4 + 2
(|r_0|^2-1)^2\nonumber\\&\;\;\;\; + 4 (1 + |r_h|)^2 (1 + |r_0|^2) +
 |r_h|^2 \big[(|r_h|-1)^2+ (1 + |r_0|)^2\big]^2\big),
\end{eqnarray}
\begin{eqnarray}                      \label{eq32}
F_{F}& = \frac{1}{8}\times\big[4(1+|r_h|)(1+|r_0||r_h|)+2(2+|r_0|+|r_h|)(2+|r_0|^2+|r_h|^2)
            \nonumber\\& \;\;\;\; +(1+|r_0|)(4
|r_h|^2-|r_h|^4+2 |r_h|^3 |r_0|+2 |r_h| |r_0|^3+|r_0|^4)\big]/
           \big(\big[(|r_h|-1)^2 \nonumber\\& \;\;\;\;
+(1+|r_0|)^2\big]\big[4+2(|r_h|-|r_0|)^2+(|r_h|^2+|r_0|^2)^2\big]+4\big[(1+|r_h|)^2
\nonumber\\&\;\;\;\; +(|r_0|-1)^2\big]
\big[8+2(|r_h|^2+|r_0|^2)^2\big]+\big[2+|r_h|(|r_h|-2)
+|r_0|(2+|r_0|)\big]\nonumber\\&\;\;\;\; \times
\big[|r_h|^8-4|r_h|^7|r_0|+4|r_h|^3|r_0|^5
 +8 |r_h|^2 |r_0|^6+4 |r_h||r_0|^7+|r_0|^8 \nonumber\\
 &\;\;\;\;-4
|r_h|^5|r_0|
 (|r_0|^2-4) +8|r_h|^6 (|r_0|^2-1)-2 |r_h|^4(4|r_0|^2+|r_0|^4-8)\big]\big),
\end{eqnarray}
\begin{eqnarray}                    \label{eq33}
\eta_{C}=\frac{(2 + |r_h|^2 + |r_0|^2)^2}{16},
\end{eqnarray}
\begin{eqnarray}                      \label{eq34}
\eta_{T}=\frac{(2 + |r_h|^2 + |r_0|^2)^2 (6 +
|r_h|^2+|r_0|^2)}{128},\;\;\;\;
\end{eqnarray}
\begin{eqnarray}                      \label{eq35}
\eta_{F}=\frac{(2 + |r_h|^2 + |r_0|^2) \big[4 + (|r_h|^2
+|r_0|^2)^2\big]
 \big[12 + (|r_h|^2 +|r_0|^2)^2\big]}{512}.
\end{eqnarray}

It is still a big challenge to achieve strong coupling in experiment
at present \cite{Hybrid2}. However, strong coupling has been
observed in the QD-cavity systems with the micropillar form
\cite{Hu6,Reithmaier,Reitzenstein,Loo} and the microdisk form
\cite{Peter,Michael}, and the QD-nanocavity systems \cite{Yoshie} in
experiment. In 2004, Reithmaier \emph{et al.} \cite{Reithmaier}
observed $g/(\kappa+\kappa_s)\simeq0.5$
[$g/(\kappa+\kappa_s)\simeq2.4$] in a $d=1.5$ $\mu$m micropillar
cavity with a quality factor of  $Q=8800$ [$Q=40000$].  In 2011, Hu
\emph{et al.} \cite{Hu4} demonstrated $g/(\kappa+\kappa_s)\simeq1.0
$ in a micropillar cavity with $\kappa_s/\kappa\simeq0.7$ and
$Q\simeq1.7\times10^4$. In 2010, Loo \emph{et al.} \cite{Loo}
reported $g=16\;\mu eV$ and $\kappa=20.5\;\mu eV$ in a $d=7.3\; \mu
m$ micropillar with $Q=65000$.

\begin{figure}
\centering \includegraphics[width=5 cm,angle=0]{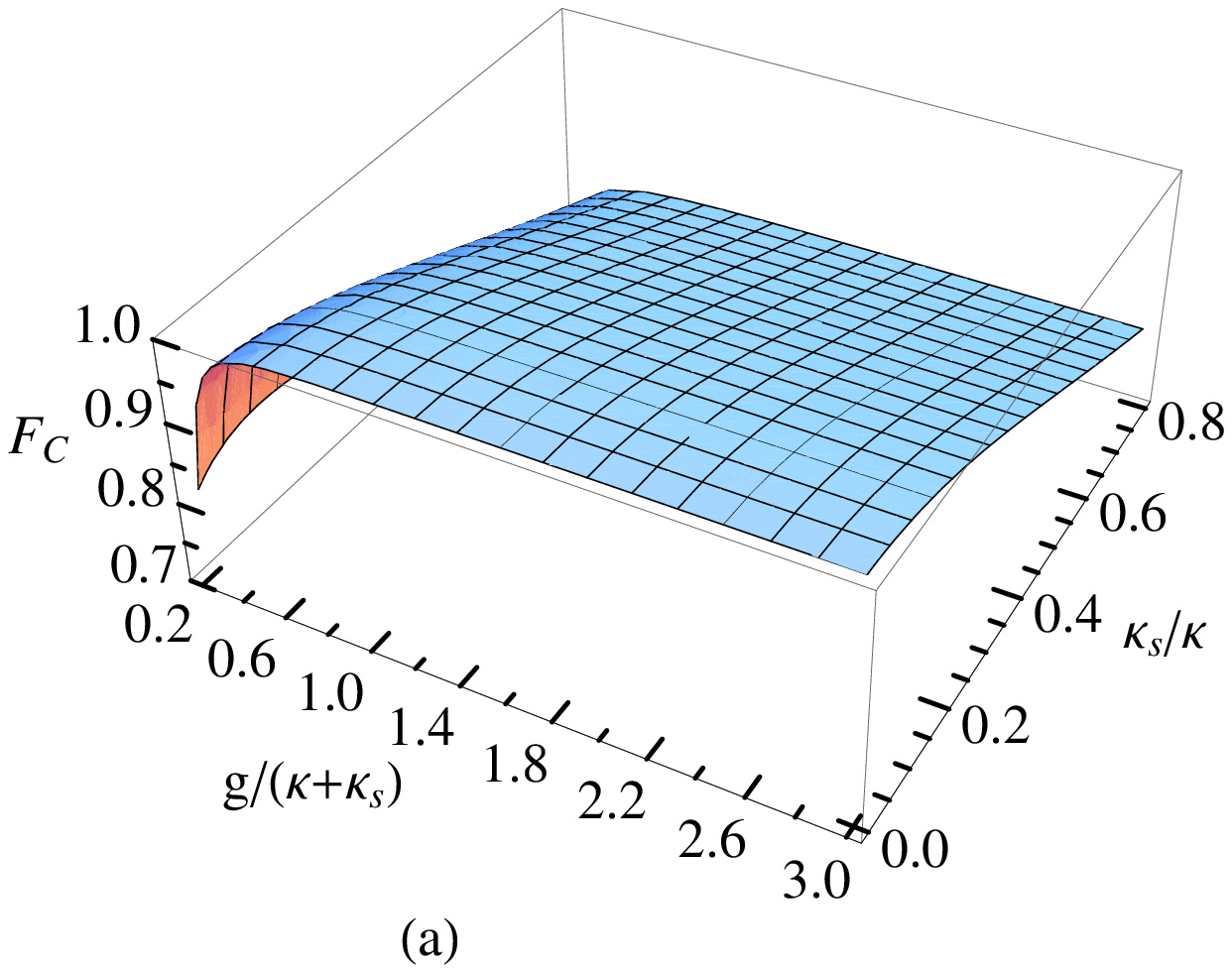}\;\;\;\;
\includegraphics[width=5 cm,angle=0]{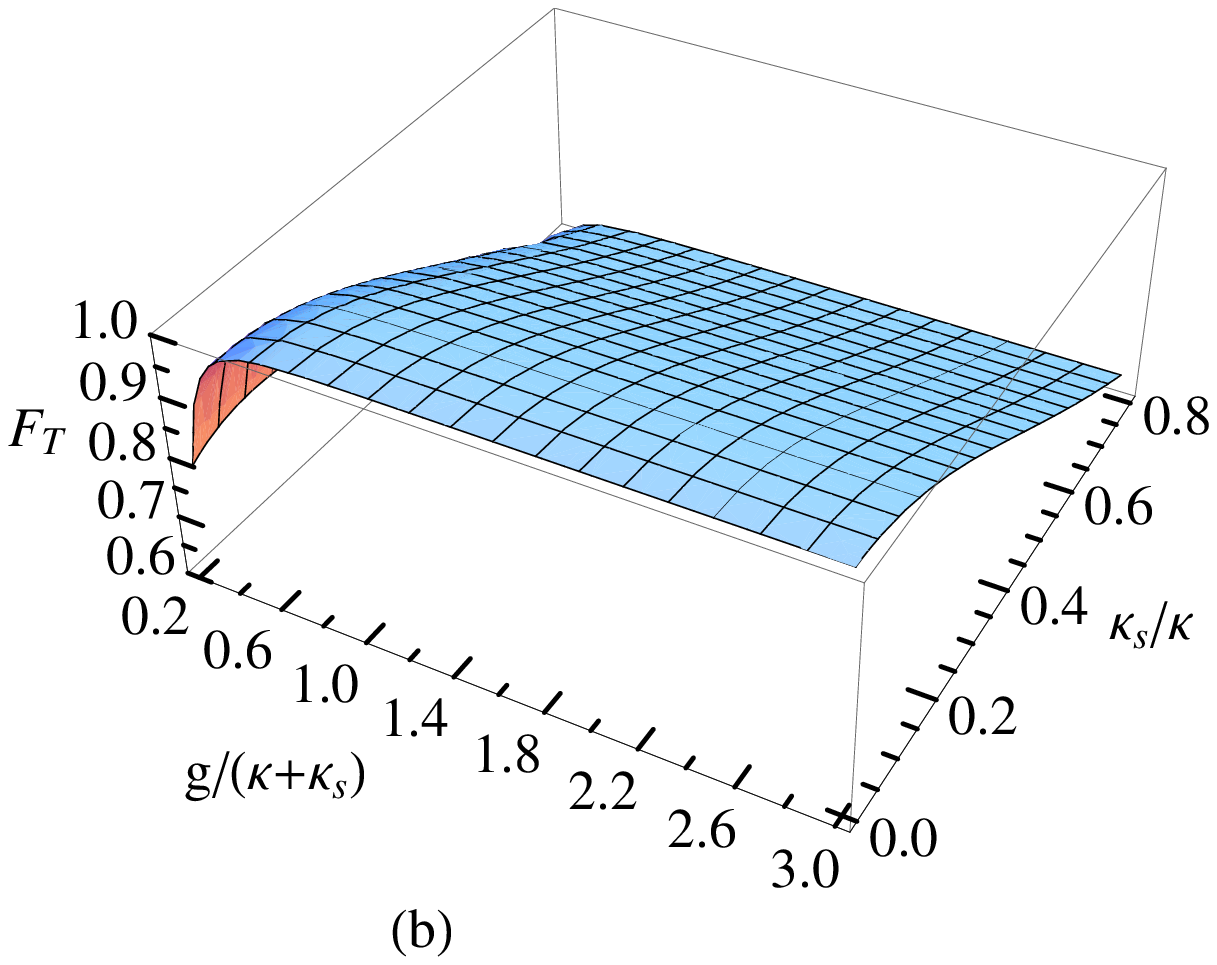}\;\;\;\;
\includegraphics[width=5 cm,angle=0]{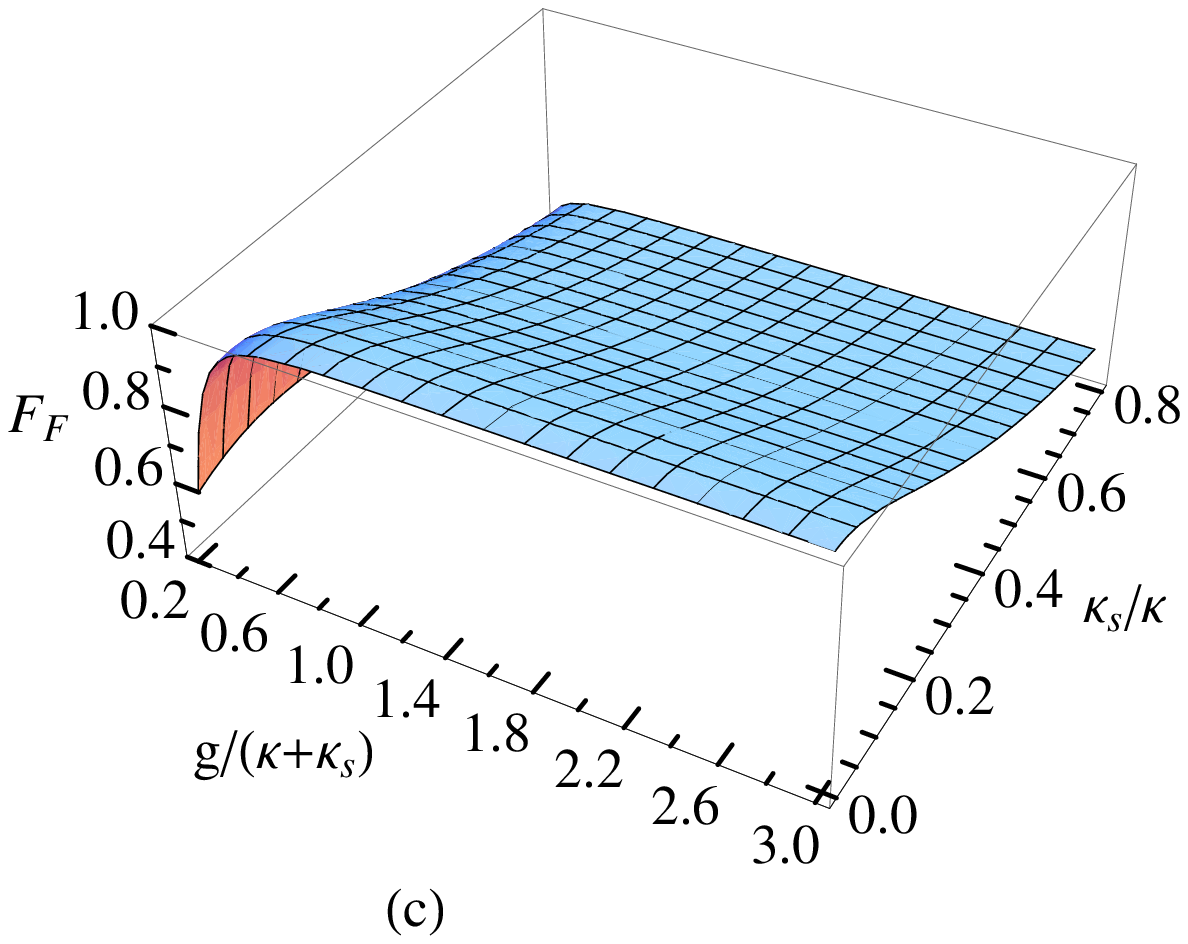}\\
\caption{ The fidelities of our universal quantum gates vs the
coupling strength $g/(\kappa+\kappa_s)$ and
 the side leakage rate  $\kappa_s/\kappa$. (a) The   fidelity of our CNOT
 gate ( $F_{C}$). (b) The   fidelity of our
Toffoli gate ($F_{T}$) . (c) The   fidelity of our Fredkin gate
($F_{F}$).  We take $\omega=\omega_c=\omega_{X^-}$ and
$\gamma/\kappa=0.1$.} \label{Fidelity}
\end{figure}

\begin{figure}          
\centering
\includegraphics[width=5.0 cm,angle=0]{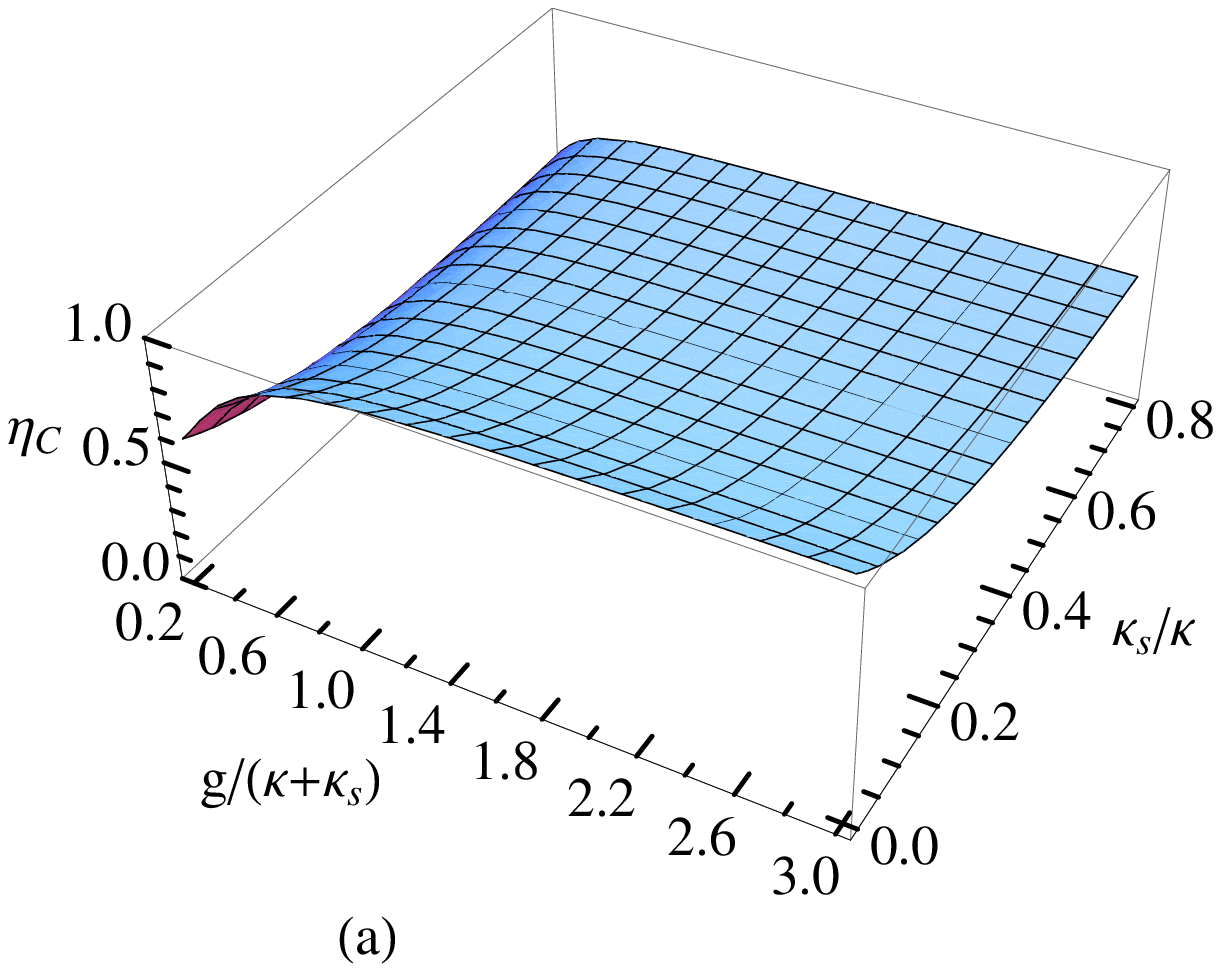}\;\;\;\;
\includegraphics[width=5.0 cm,angle=0]{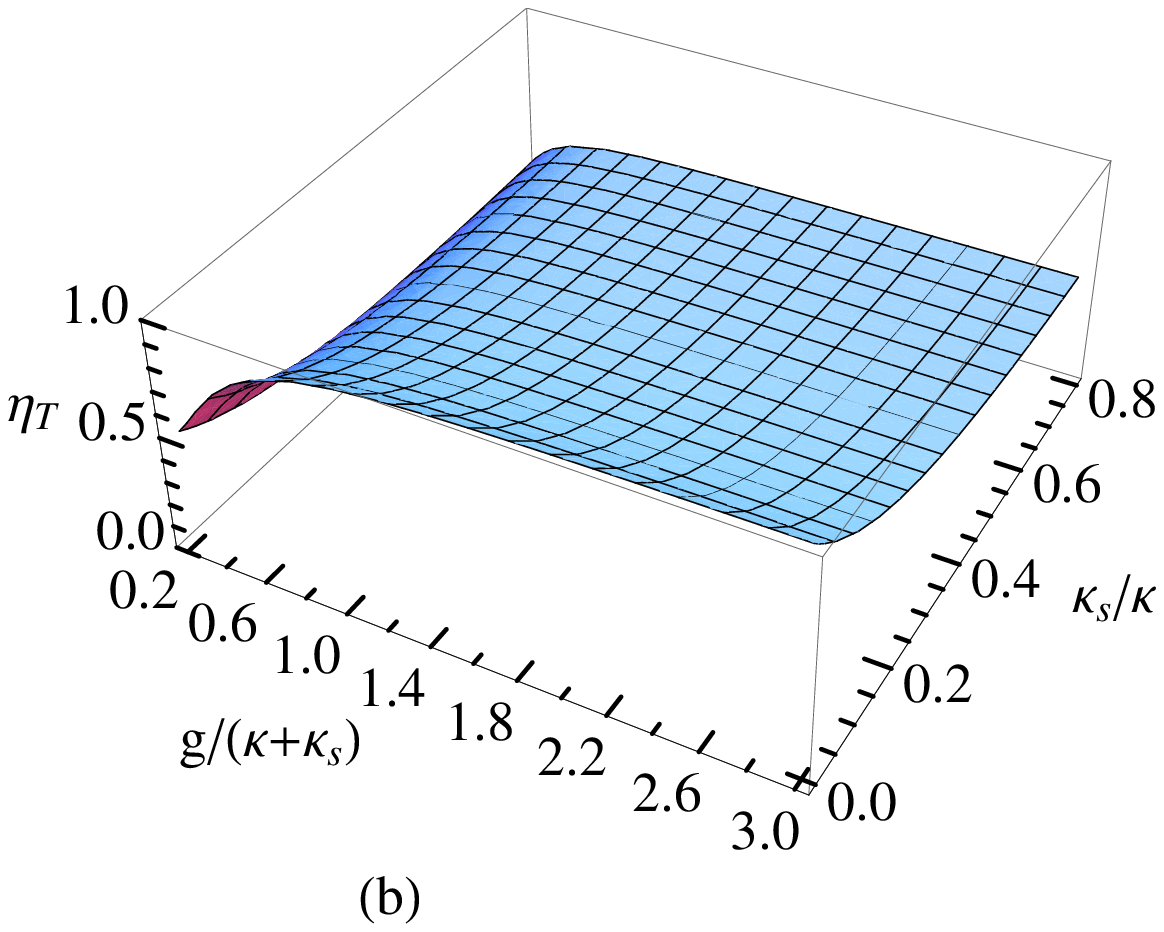}\;\;\;\;
\includegraphics[width=5.0 cm,angle=0]{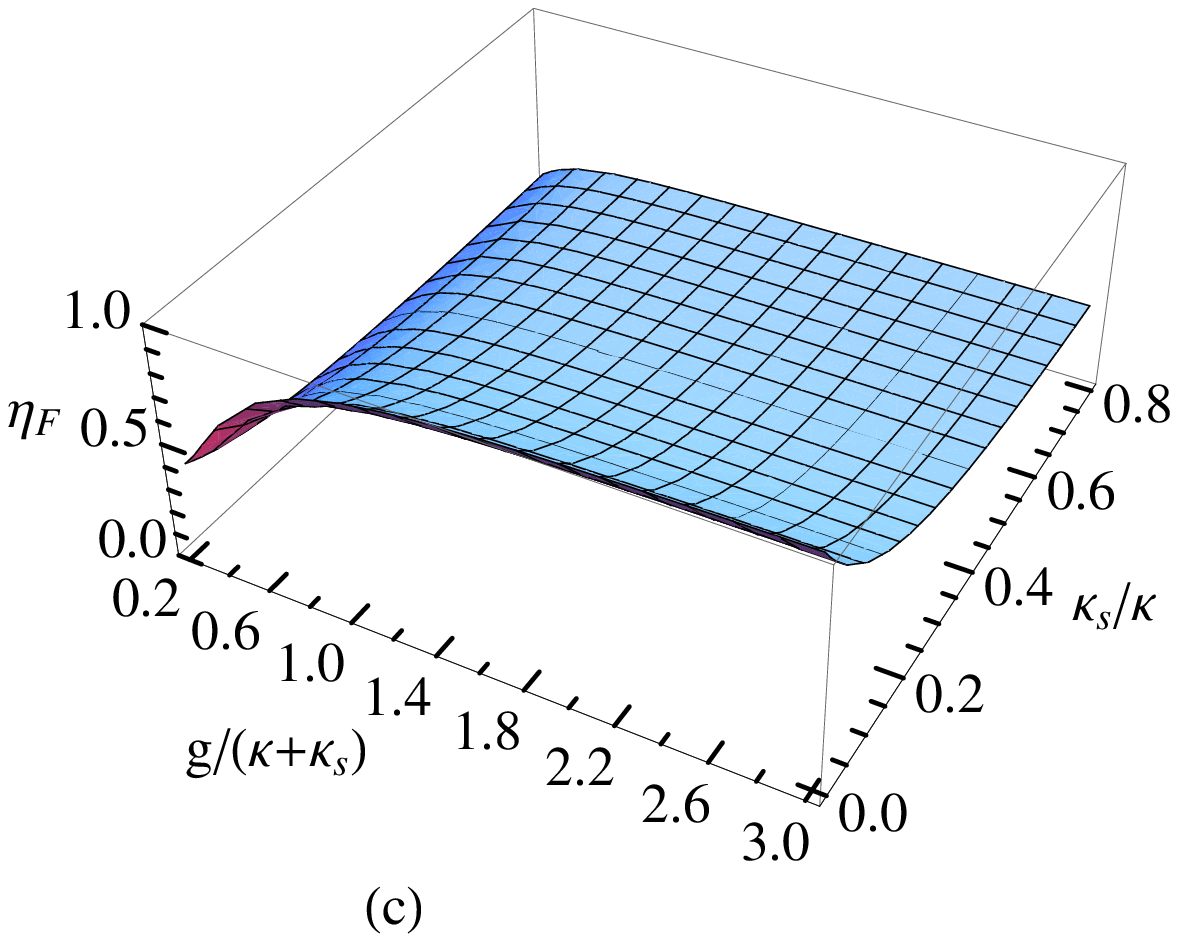}\\
\caption{ The efficiencies of our universal quantum gates vs the
coupling strength $g/(\kappa+\kappa_s)$ and
 the side leakage rate  $\kappa_s/\kappa$. (a) The efficiency of our CNOT
 gate ($\eta_{C}$). (b) The efficiency of our
Toffoli gate ($\eta_{T}$) . (c) The efficiency of our Fredkin gate
($\eta_{F}$). We take $\omega=\omega_c=\omega_{X^-}$  and
$\gamma/\kappa=0.1$.} \label{Efficiency}
\end{figure}

The fidelities and the efficiencies of our universal quantum gates,
which  vary  with the coupling strength and the side leakage rate,
are shown in Figs. \ref{Fidelity} and \ref{Efficiency},
respectively. From these figures, one can see that our schemes are
feasible in both the strong coupling regime and the weak coupling
regime. $\kappa_s$ can be made rather small by improving the sample
growth or the etching process.

A QD system  has the discrete atom-like energy levels and a spectrum
of the ultra-narrow transition that is tunable with the size of the
quantum dot. The growth techniques of QDs  produce the size
variations of the QDs. The spectral line-width inhomogeneous
broadening is caused by the fluctuations in the size and shape of a
QD, and it has gained the widespread attention \cite{inhomg2}. The
spectral inhomogeneity is an important property and it is not
necessarily a negative consequence for their applications in quantum
information processing. The imperfect QD in a realistic system,
i.e., the shape of the sample and the strain field distribution are
not symmetric, reduces the fidelities of the gates and it can be
decreased by designing the shape and the size of the sample or
encoding the qubits on a different type of QDs \cite{Hu2,Hu4}.

The information between the photon medium and the QD spins is
transferred by the exciton.  That is,  the exciton dephasing reduces
the fidelities of the gates. The exciton dephasing, including the
optical dephasing and the spin dephasing, is sensitive to the dipole
coherence time  $T_2$ and the cavity-photon coherence time $\tau$.
The exciton dephasing reduces the fidelities of the universal
quantum gates less than 10\% as it reduces the fidelities by a
factor
\begin{equation}
 1-\exp(-\tau/T_2),
\end{equation}
and the ultralong optical coherence time of the dipole $T_2$ can
reach several picoseconds at a low temperature
\cite{opticldeph1,opticldeph2}, while the cavity-photon coherence
time $\tau$ is around 10 picoseconds in a InGaAs QD. The QD-hole
spin coherence time $T_2$ is long more than 100 nanoseconds
\cite{Hole-QD}.

\section{Discussion and summary}\label{sec6}

Quantum logic gates are essential building blocks in quantum
computing and quantum information processing \cite{book}. CNOT gates
are used widely in quantum computing. Directly physical realization
of multiqubit gates is a main direction as the optimal length of the
unconstructed circuit for a generic $n$-qubit gate is
$[(4^n-3n-1)/4]$ \cite{3CNOT4}.

Some significant progress has been made in  realizing universal
quantum gates.  Refs. \cite{Hybrid1,Hybrid2,NV2} present some
interesting schemes for the  quantum gates on hybrid light-matter or
electron-nuclear qubits. Based on parity-check gates, the CNOT gate
on  moving electron qubits is proposed in 2004, assisted by an
additional electron qubit \cite{MoveCNOT}. A Toffoli gate on atom
qubits with a success probability of 1/2 is constructed by Wei
\emph{et al.} in 2008 \cite{Toffoliwei}. Our CNOT, Toffoli, and
Fredkin gates are compact, simple, and economic as  the ancilla
qubits, employed in \cite{photon5,photon9,MoveCNOT}, are not
required, and only a single-photon medium is employed. The proposals
for the Toffoli  and Fredkin gates beat their synthesis with
two-qubit entangling gates and single-qubit gates largely. The
optimal synthesis of a three-qubit Toffoli gate requires six CNOT
gates \cite{Toffolicost} and five  quantum entangling gates on two
individual qubits are required to synthesize a three-qubit Fredkin
gate \cite{Fredkincost}. All our schemes are deterministic and the
qubits for the gates are stationary. The side leakage rate of a
single-side cavity is usually lower than  that of a double-side one
\cite{Hu3}. Moreover, a QD is easier to be confined in a  cavity
than an atom \cite{hard,atom4}.

In summary, we have proposed some compact schemes for implementing
quantum computing on solid-state electron-spin qubits in the QDs
assisted by single-side resonant optical microcavities in a
deterministic way. Based on the fact that the $R$-polarized and the
$L$-polarized photons reflected by the QD-cavity contribute
different phase shifts, the compact quantum circuits for the CNOT,
Toffoli, and  Fredkin gates on  the stationary electron-spin qubits
are achieved by some input-output processes of a single-photon
medium and some classical feed-forward operations. Our proposals are
compact and economic as the additional QD-spin qubits are not
required and our schemes for implementing the multiqubit gates beat
their synthesis with two-qubit entangling gates and single-qubit
gates largely. The success probabilities of our universal quantum
gates are 100\% in principle. With current  technology, our schemes
are feasible. Together with single-qubit gates, our universal
quantum gates are sufficient for any quantum computing in
solid-state QD-spin systems.

\section*{Acknowledgments}

This work is supported by the National Natural Science Foundation of
China under Grant No. 11174039,  NECT-11-0031, and the Open
Foundation of State key Laboratory of Networking and Switching
Technology (Beijing University of Posts and Telecommunications)
under Grant No. SKLNST-2013-1-13.


\end{document}